\documentclass[preprintnumbers,amsmath,amssymbm,prl]{revtex4}
\usepackage{epsfig}
\usepackage{graphicx}

\begin{document}
\title{No-bomb theorem for charged Reissner-Nordstr\"om black holes}
\author{Shahar Hod}
\affiliation{The Ruppin Academic Center, Emeq Hefer 40250, Israel}
\affiliation{ } \affiliation{The Hadassah Institute, Jerusalem
91010, Israel}
\date{\today}

\begin{abstract}
\ \ \ The fundamental role played by black holes in many areas of
physics makes it highly important to explore the nature of their
stability. The stability of charged Reissner-Nordstr\"om black holes
to {\it neutral} (gravitational and electromagnetic) perturbations
was established almost four decades ago. However, the stability of
these charged black holes under {\it charged} perturbations has
remained an open question due to the complexity introduced by the
well-known phenomena of superradiant scattering: A charged scalar
field impinging on a charged Reissner-Nordstr\"om black hole can be
{\it amplified} as it scatters off the hole. If the incident field
has a non-zero rest mass, then the mass term effectively works as a
mirror, preventing the energy extracted from the hole from escaping
to infinity. One may suspect that the superradiant amplification of
charged fields by the charged black hole may lead to an instability
of the Reissner-Nordstr\"om spacetime (in as much the same way that
rotating Kerr black holes are unstable under rotating scalar
perturbations). However, in this Letter we show that, for charged
Reissner-Nordstr\"om black holes in the regime ${(Q/M)}^2\leq 8/9$,
the two conditions which are required in order to trigger a possible
superradiant instability [namely: (1) the existence of a trapping
potential well outside the black hole, and (2) superradiant
amplification of the trapped modes] cannot be satisfied
simultaneously. Our results thus support the stability of charged
Reissner-Nordstr\"om black holes under charged scalar perturbations
in the regime ${(Q/M)}^2\leq 8/9$.
\end{abstract}
\bigskip
\maketitle


It is well-known that rotating Kerr black holes suffer from
superradiant instability which may be triggered by massive scalar
perturbations \cite{PressTeu1,Teu,Hodp1}. Superradiant scattering is
a well-known phenomena in quantum systems \cite{Mano,Grein} as well
as in classical ones \cite{Zel,Car}. Considering a wave of the form
$e^{im\phi}e^{-i\omega t}$ incident upon a rotating object whose
angular velocity is $\Omega$, one finds that if the frequency
$\omega$ of the incident wave satisfies the relation
\begin{equation}\label{Eq1}
\omega<m\Omega\  ,
\end{equation}
then the scattered wave is amplified.

A bosonic field impinging upon a rotating Kerr black hole can be
amplified if the superradiance condition (\ref{Eq1}) is satisfied,
where in this case $\Omega$ is the angular velocity of the
black-hole horizon. Press and Teukolsky suggested to use this
mechanism to build a black-hole bomb \cite{PressTeu2}: If one
surrounds the black hole by a reflecting mirror, the wave will
bounce back and forth between the black hole and the mirror
amplifying itself each time. It was later realized
\cite{Dam,Zour,Det,Furu,Dolan,CarYo,HodHod,Beyer2,HodN} that a
natural mirror exists in a system composed of a rotating black hole
and a {\it massive} scalar field: In this case the mass term
effectively works as a mirror for modes in the regime
$\omega<\mu\equiv {\cal M}G/\hbar c$, where ${\cal M}$ is the mass
of the field. The gravitational force binds the massive field and
keeps it from escaping to infinity. At the event horizon some of the
field goes down the black hole, and if the frequency of the wave is
in the superradiance regime (\ref{Eq1}) then the field is amplified.
In this way the field is amplified at the horizon while being bound
away from infinity. As a consequence, the rotational energy
extracted from the black hole by the incident bosonic field grows
exponentially over time
\cite{Dam,Zour,Det,Furu,Dolan,CarYo,HodHod,Beyer2,HodN}.

A similar amplification of waves occurs when a {\it charged} bosonic
field impinges upon a {\it charged} Reissner-Nordstr\"om black hole
\cite{Bekch}. In this case the superradiant scattering occurs for
frequencies in the regime
\begin{equation}\label{Eq2}
\omega<q\Phi\  ,
\end{equation}
where $q$ is the charge coupling constant of the field and
$\Phi=Q/r_+$ is the electric potential of the charged black hole.
(Here $r_+$ and $Q$ \cite{NoteQ} are the horizon radius and electric
charge of the black hole, respectively). The superradiant scattering
of charged scalar waves in the regime (\ref{Eq2}) results in the
extraction of Coulomb energy and electric charge from the charged
black hole \cite{Bekch}. One may suspect that the amplification of
charged massive fields by charged black holes may lead to an
instability of the Reissner-Nordstr\"om spacetime (in as much the
same way that the amplification of rotating massive fields by
rotating black holes leads to an instability of the Kerr spacetime).

The physical interest in Reissner-Nordstr\"om (RN) black holes is
mainly motivated by the fact that these charged black holes share
many common features with the physically more relevant rotating Kerr
black holes. In particular, the global spacetime structures of
rotating Kerr black holes and charged RN black holes are almost
identical \cite{HodPirmass}. It is therefore of physical interest to
explore the stability (or instability) regime of these charged black
holes.

The stability of RN black holes under {\it neutral} (gravitational
and electromagnetic) perturbations was established long ago by
Moncrief \cite{Monf1,Monf2}. However, much less is known about the
stability of these charged black holes under {\it charged}
perturbation fields \cite{Hodn}. It should be realized that such
charged perturbation fields are expected to be an important
ingredient of charged gravitational collapse to form a charged RN
black hole \cite{HodPirpam}. To the best of our knowledge, no
stability proof exists in the literature for generic charged RN
black holes coupled to {\it charged} perturbation fields. As a
matter of fact, the well-known superradiant instability of rotating
Kerr black holes due to massive scalar perturbations [in the regime
(\ref{Eq1})] may lead one to suspect that an analogous instability
may occur for charged RN black holes -- this time due to
superradiant amplification of {\it charged} scalar waves [in the
regime (\ref{Eq2})].

The main goal of the present Letter is to provide evidence for the
stability of charged RN black holes under charged scalar
perturbations. In particular, we shall show below that, for charged
RN black holes in the regime ${(Q/M)}^2\leq 8/9$, the two conditions
which are required in order to trigger the superradiant instability
[namely: (1) the existence of a trapping potential well outside the
black hole, and (2) superradiant amplification of the trapped modes]
cannot be satisfied simultaneously. Our results thus support the
stability of these charged black holes.

The physical system we consider consists of a massive charged scalar
field coupled to a Reissner-Nordstr\"om black hole of mass $M$ and
electric charge $Q$. The black-hole spacetime is described by the
line element
\begin{equation}\label{Eq3}
ds^2=-f(r)dt^2+{1\over{f(r)}}dr^2+r^2(d\theta^2+\sin^2\theta
d\phi^2)\ ,
\end{equation}
where
\begin{equation}\label{Eq4}
f(r)\equiv 1-{{2M}\over{r}}+{{Q^2}\over{r^2}}\  .
\end{equation}
Here $r$ is the Schwarzschild areal coordinate. (We use natural
units in which $G=c=\hbar=1$.)

The dynamics of the charged massive scalar field $\Psi$ in the RN
spacetime is governed by the Klein-Gordon equation
\cite{Stro,HodPirpam,HodCQG2,Hodpla}
\begin{equation}\label{Eq5}
[(\nabla^\nu-iqA^\nu)(\nabla_{\nu}-iqA_{\nu}) -\mu^2]\Psi=0\  ,
\end{equation}
where $A_{\nu}=-\delta_{\nu}^{0}{Q/r}$ is the electromagnetic
potential of the black hole. Here $q$ and $\mu$ are the charge and
mass of the field, respectively. [Note that $q$ and $\mu$ stand for
$q/\hbar$ and $\mu/\hbar$, respectively. Thus, they have the
dimensions of $($length$)^{-1}$.] One may decompose the field as
\begin{equation}\label{Eq6}
\Psi_{lm}(t,r,\theta,\phi)=e^{im\phi}S_{lm}(\theta)R_{lm}(r)e^{-i\omega
t}\ ,
\end{equation}
where $\omega$ is the conserved frequency of the mode, $l$ is the
spherical harmonic index, and $m$ is the azimuthal harmonic index
with $-l\leq m\leq l$. (We shall henceforth omit the indices $l$ and
$m$ for brevity.) The sign of $\omega_I$ determines whether the
solution is decaying $(\omega_I<0)$ or growing $(\omega_I>0)$ in
time. Remembering that any instability must set in via a
real-frequency mode \cite{Dam,Zour,Hartle} we shall consider here
modes with $|\omega_I|\ll \omega_R$.

With the decomposition (\ref{Eq6}), $R$ and $S$ obey radial and
angular equations both of confluent Heun type coupled by a
separation constant $K_l=l(l+1)$ \cite{Heun,Flam,Abram}. The radial
Klein-Gordon equation is given by \cite{Stro,HodPirpam,HodCQG2}
\begin{equation}\label{Eq7}
\Delta{{d} \over{dr}}\Big(\Delta{{dR}\over{dr}}\Big)+UR=0\ ,
\end{equation}
where
\begin{equation}\label{Eq8}
\Delta\equiv r^2-2Mr+Q^2\  ,
\end{equation}
and
\begin{equation}\label{Eq9}
U\equiv(\omega r^2-qQr)^2 -\Delta[\mu^2r^2+l(l+1)]\  .
\end{equation}
The zeros of $\Delta$,
\begin{equation}\label{Eq10}
r_{\pm}=M\pm (M^2-Q^2)^{1/2}\  ,
\end{equation}
are the black-hole (event and inner) horizons.

We are interested in solutions of the radial equation (\ref{Eq7})
with the physical boundary conditions of purely ingoing waves at the
black-hole horizon (as measured by a local observer) and a decaying
(bounded) solution at spatial infinity \cite{Zour,Dolan}. That is,
\begin{equation}\label{Eq11}
R \sim e^{-i (\omega-qQ/r_+)y}\ \ \text{ as }\ r\rightarrow r_+\ \
(y\rightarrow -\infty)\ ,
\end{equation}
and
\begin{equation}\label{Eq12}
R \sim y^{-iqQ}e^{-\sqrt{\mu^2-\omega^2}y}\ \ \text{ as }\
r\rightarrow\infty\ \ (y\rightarrow \infty)\  ,
\end{equation}
where the ``tortoise" radial coordinate $y$ is defined by
$dy=(r^2/\Delta)dr$. For frequencies in the superradiant regime
(\ref{Eq2}), the boundary condition (\ref{Eq11}) describes an
outgoing flux of energy and charge from the charged black hole
\cite{Dam,Bekch}. Note also that a bound state (a state decaying
exponentially at spatial infinity) is characterized by
\begin{equation}\label{Eq13}
\omega^2<\mu^2\  .
\end{equation}
The boundary conditions (\ref{Eq11})-(\ref{Eq13}) single out a
discrete set of resonances $\{\omega_n\}$ which correspond to the
bound states of the charged massive field \cite{Dam}. (We note that,
in addition to the bound states of the charged field, the field also
has an infinite set of discrete quasinormal resonances
\cite{HodCQG2,Hodpla,Will} which are characterized by outgoing waves
at spatial infinity.)

It is convenient to define a new radial function $\psi$ by
\begin{equation}\label{Eq14}
\psi\equiv \Delta^{1/2}R\  ,
\end{equation}
in terms of which the radial equation (\ref{Eq7}) can be written in
the form of a Schr\"odinger-like wave equation
\begin{equation}\label{Eq15}
{{d^2\psi}\over{dr^2}}+(\omega^2-V)\psi=0\  ,
\end{equation}
where
\begin{equation}\label{Eq16}
\omega^2-V={{U+M^2-Q^2}\over{\Delta^2}}\  .
\end{equation}

As discussed above, two ingredients are required in order to trigger
a superradiant instability:
\begin{itemize}
\item{The existence of a trapping potential well (supporting meta-stable bound states) outside the black hole. Moreover,
in order to support these meta-stable bound states, the well itself
must be separated from the black-hole horizon by a potential
barrier.}
\item{The existence of superradiant amplification of the charged fields.}
\end{itemize}
We shall now show that these two conditions cannot be satisfied
simultaneously for charged RN black holes in the regime
${(Q/M)}^2\leq 8/9$. In particular, we shall prove that there are no
meta-stable bound states of the charged field (i.e., there is no
potential well in the black-hole exterior which is separated from
the horizon by a potential barrier) in the superradiant regime
(\ref{Eq2}). To that end, we shall analyze the behavior of the
effective potential $V(r;M,Q,\mu,q,\omega,l)$. [It should be
realized that such an analysis is not an easy task since the
effective potential $V$ is characterized by a set of six parameters:
$\{M,Q,\mu,q,\omega,l\}$.]

The derivative of the effective potential is given by the rational
function
\begin{eqnarray}\label{Eq17}
V'(r;M,Q,\mu,q,\omega,l)&=&-{{2}\over{\Delta^3}}\Big[(M\mu^2+Qq\omega-2M\omega^2)r^4
+[-2M^2\mu^2-Q^2(q^2+\mu^2)+2MQq\omega+2Q^2\omega^2+l(l+1)]r^3\nonumber\\&&
+[3MQ^2\mu^2-3Q^3q\omega-3Ml(l+1)]r^2+[Q^4(q^2-\mu^2)-2(M^2-Q^2)+(2M^2+Q^2)l(l+1)]r\nonumber\\&&
+2M(M^2-Q^2)-MQ^2l(l+1)\Big]\ .
\end{eqnarray}
It is convenient to define a new variable
\begin{equation}\label{Eq18}
z\equiv r-r_-\  ,
\end{equation}
in terms of which Eq. (\ref{Eq17}) can be written as a $4$th order
polynomial function
\begin{equation}\label{Eq19}
-{{\Delta^3}\over{2}}V'(r;M,Q,\mu,q,\omega,l)=az^4+bz^3+cz^2+dz+e\ ,
\end{equation}
where the coefficients $\{a,b,c,d,e\}$ are given by
\begin{equation}\label{Eq20}
a=M\mu^2+Qq\omega-2M\omega^2\  ,
\end{equation}
\begin{equation}\label{Eq21}
b=(4Mr_--2M^2-Q^2)\mu^2+(-8Mr_-+2Q^2)\omega^2+2Q(M+2r_-)q\omega-Q^2q^2+l(l+1)\
,
\end{equation}
\begin{equation}\label{Eq22}
c=-3r^2_-(r_+-M)\mu^2+3r^3_-({{qQ}\over{r_-}}-\omega)(2\omega-{{qQ}\over{r_-}})-3(r_+-M)l(l+1)\
,
\end{equation}
\begin{equation}\label{Eq23}
d=2r^2_-(M^2-Q^2)\mu^2+Q^2r_-(r_+-3r_-)q^2+2r^3_-(3r_+-4M)\omega^2+2Qr^2_-(2r_-+3M-3r_+)q\omega
+2(M^2-Q^2)[l(l+1)-1]\ ,
\end{equation}
and
\begin{equation}\label{Eq24}
e=2r^4_-(r_+-M)(\omega-{{qQ}\over{r_-}})^2+2(r_+-M)^3\ .
\end{equation}

We denote the roots of $V'(z)=0$ by $\{z_1,z_2,z_3,z_4\}$.
Below we shall analyze the properties (and, in particular, the
signs) of these roots. To that end, we shall use the well-known
relations
\begin{equation}\label{Eq25}
z_1+z_2+z_3+z_4=-{b\over a}\  ,
\end{equation}
\begin{equation}\label{Eq26}
z_1\cdot z_2+z_1\cdot z_3+z_1\cdot z_4+z_2\cdot z_3+z_2\cdot
z_4+z_3\cdot z_4={c\over a}\ ,
\end{equation}
\begin{equation}\label{Eq27}
z_1\cdot z_2\cdot z_3+z_1\cdot z_2\cdot z_4+z_1\cdot z_3\cdot
z_4+z_2\cdot z_3\cdot z_4=-{d\over a}\  ,
\end{equation}
and
\begin{equation}\label{Eq28}
z_1\cdot z_2\cdot z_3\cdot z_4={e\over a}\  ,
\end{equation}
for the four solutions of a quartic equation.

The superradiant condition (\ref{Eq2}) and the bound state condition
(\ref{Eq13}) imply
\begin{equation}\label{Eq29}
0\leq\omega<\text{min}\{{{qQ}/{r_+}},\mu\}\  .
\end{equation}
Note that the dependence of the coefficient $a$ on $\omega$ is in
the form of a convex parabola, see Eq. (\ref{Eq20}). Thus,
$a(\omega)$ is minimized at the edges of the interval (\ref{Eq29}).
Substituting $\omega=0$ into (\ref{Eq20}), one finds $a=M\mu^2>0$.
Substituting $\omega\to qQ/r_+$ for $qQ/r_+\leq\mu$, one finds
$a=M\mu^2+{{Q^2q^2}\over{r_+}}(1-{{2M}\over{r_+}})>0$. Substituting
$\omega\to\mu$ for $\mu<qQ/r_+$, one finds
$a=\mu(Qq-M\mu)>\mu^2(r_+-M)\geq 0$. We thus conclude that
\begin{equation}\label{Eq30}
a>0\
\end{equation}
in the entire parameter space. This implies that [see Eq.
(\ref{Eq19})]
\begin{equation}\label{Eq31}
V'(r\to\infty)\to 0^-\
\end{equation}
in the superradiant regime. Note also that
\begin{equation}\label{Eq32}
V(r\to r_+)\to -\infty\  ,
\end{equation}
and
\begin{equation}\label{Eq33}
V(r\to r_-)\to -\infty\  .
\end{equation}
Equations (\ref{Eq31}) and (\ref{Eq32}) imply that $V(r)$ has at
least one maximum point in the physical region $r>r_+$. We denote it
by $z_4$, where
\begin{equation}\label{Eq34}
z_4>0\  .
\end{equation}
Equations (\ref{Eq32})-(\ref{Eq33}) imply that $V(r)$ has at least
one maximum point in the non-physical interval $r_-<r<r_+$. We
denote it by $z_3$, where
\begin{equation}\label{Eq35}
z_4>z_3>0\  .
\end{equation}

In addition, it is obvious from (\ref{Eq24}) that
\begin{equation}\label{Eq36}
e>0\  .
\end{equation}
Equations (\ref{Eq28}), (\ref{Eq30}), and (\ref{Eq36}) imply that
\begin{equation}\label{Eq37}
z_1\cdot z_2\cdot z_3\cdot z_4>0\  .
\end{equation}

We shall now focus on charged RN black holes in the regime
\begin{equation}\label{Eq38}
{(Q/M)}^2\leq {8\over 9}\  .
\end{equation}
In this case one finds $qQ/r_+\leq qQ/2r_-$, which implies [see Eq.
(\ref{Eq29})] $0\leq\omega<qQ/2r_-$. Substituting this inequality
into Eq. (\ref{Eq22}), one deduces that
\begin{equation}\label{Eq39}
c<0\  .
\end{equation}
Equations (\ref{Eq26}), (\ref{Eq30}), and (\ref{Eq39}) imply that
\begin{equation}\label{Eq40}
z_1\cdot z_2+z_1\cdot z_3+z_1\cdot z_4+z_2\cdot z_3+z_2\cdot
z_4+z_3\cdot z_4<0\ .
\end{equation}
Taking cognizance of Eqs. (\ref{Eq37}) and (\ref{Eq40}), one learns
that there are two negative roots (we denote them by $z_1$ and
$z_2$, where $z_1\leq z_2<0$) and two positive roots ($0<z_3<z_4$)
of $V'(z)$.

We have therefore proved that, for bound states (with $\omega<\mu$)
in the superradiant regime (with $\omega<qQ/r_+$), there are two
negative roots and two positive roots of $V'(z)$ in the entire
parameter space (characterized by the six parameters $\{M,Q\leq
\sqrt{8/9}M,\mu,q,\omega,l\}$). The negative roots $z_1$ and $z_2$
correspond to two roots of $V'(r)$ in the non-physical region
$r<r_-$. The positive root $z_3$ corresponds to a maximum point of
$V(r)$ in the non-physical region $r_-<r<r_+$. The physical root
$z_4>0$ (or equivalently, $r_4>r_+$) corresponds to a {\it maximum}
point of $V(r)$ outside the horizon.

In conclusion, we have shown that in the superradiant regime $V(r)$
has only one maximum (and {\it no} minima at all) outside the event
horizon. This implies that there is no potential well in the
black-hole exterior which is separated from the horizon by a
potential barrier. Thus, there are no meta-stable bound states of
the charged massive scalar field in the superradiant regime.


In summary, motivated by the well-known phenomena of superradiant
instability of rotating Kerr black holes due to massive scalar
perturbations, we have investigated here the possible existence of
an analogous instability for charged Reissner-Nordstr\"om black
holes due to charged massive perturbations. We have proved that the
two conditions which are necessary for the existence of a
superradiant instability [namely: (1) the existence of a trapping
potential well, and (2) superradiant amplification of the trapped
modes] cannot be satisfied simultaneously for charged
Reissner-Nordstr\"om black holes in the regime $Q/M\leq \sqrt{8/9}$.
That is, we have shown that in the exterior of these charged black
holes there are no meta-stable bound states of the charged scalar
fields in the superradiant regime. Thus, the dynamics of charged
massive scalar fields in the Reissner-Nordstr\"om spacetime with
$Q/M\leq \sqrt{8/9}$ is expected to be stable.

\bigskip
\noindent
{\bf ACKNOWLEDGMENTS}
\bigskip

This research is supported by the Carmel Science Foundation. I thank
Oded Hod, Yael Oren, Arbel M. Ongo and Ayelet B. Lata for helpful
discussions.



\begin{thebibliography}{99}

\bibitem{PressTeu1} W. H. Press and S. A. Teukolsky, Astrophys. J. {\bf 185}, 649 (1973).

\bibitem{Teu} S. A. Teukolsky, Phys. Rev. Lett. {\bf 29}, 1114 (1972);
Astrophys. J. {\bf 185}, 635 (1973).

\bibitem{Hodp1} S. Hod, Phys. Rev. D {\bf 58}, 104022 (1998) [arXiv:gr-qc/9811032]; S. Hod,
Phys. Rev. D {\bf 61}, 024033 (2000) [arXiv:gr-qc/9902072]; S. Hod,
Phys. Rev. D {\bf 61}, 064018 (2000) [arXiv:gr-qc/9902073]; L.
Barack, Phys. Rev. D {\bf 61}, 024026 (2000); S. Hod, Phys. Rev.
Lett. {\bf 84}, 10 (2000) [arXiv:gr-qc/9907096]; R. J. Gleiser, R.
H. Price, and J. Pullin, Class. Quant. Grav. {\bf 25}, 072001
(2008); M. Tiglio, L. E. Kidder, and S. A. Teukolsky, Class. Quant.
Grav. {\bf 25}, 105022 (2008); S. Hod, Phys. Lett. B {\bf 666}, 483
(2008) [arXiv:0810.5419]; S. Hod, Phys. Rev. D 78, 084035 (2008)
[arXiv:0811.3806]; A. Zenginoglu and M. Tiglio, Phys. Rev.D {\bf
80}, 024044 (2009); A. J. Amsel, G.  T. Horowitz, D. Marolf, and M.
M. Roberts, J. High Energy Phys. 0909:044 (2009).

\bibitem{Mano} C. A. Manogue, Annals of Physics {\bf 181}, 261
(1988).

\bibitem{Grein} W. Greiner, B. M\"uller and J. Rafelski, {\it
Quantum electrodynamics of strong fields}, (Springer-Verlag, Berlin,
1985).

\bibitem{Zel} Ya. B. Zel`dovich, Pis`ma Zh. Eksp. Teor. Fiz. {\bf
14}, 270 (1971) [JETP Lett. {\bf 14}, 180 (1971)]; Zh. Eksp. Teor.
Fiz. {\bf 62}, 2076 (1972) [Sov. Phys. JETP {\bf 35}, 1085 (1972)].

\bibitem{Car} V. Cardoso, O. J. C. Dias, J. P. S. Lemos and S.
Yoshida, Phys. Rev. D {\bf 70}, 044039 (2004); Erratum-ibid. D {\bf
70}, 049903 (2004); V. Cardoso and J. P. S. Lemos, Phys. Lett. B
{\bf 621}, 219 (2005).

\bibitem{PressTeu2} W. H. Press and S. A. Teukolsky, Nature {\bf
238}, 211 (1972).

\bibitem{Dam} T. Damour, N. Deruelle and R. Ruffini, Lett. Nuovo
Cimento {\bf 15}, 257 (1976).

\bibitem{Zour} T. M. Zouros and D. M. Eardley, Annals of physics
{\bf 118}, 139 (1979).

\bibitem{Det} S. Detweiler, Phys. Rev. D {\bf 22}, 2323 (1980).

\bibitem{Furu} H. Furuhashi and Y. Nambu, Prog. Theor. Phys. {\bf 112}, 983
(2004).

\bibitem{Dolan} S. R. Dolan, Phys. Rev. D {\bf 76}, 084001 (2007).

\bibitem{CarYo}  V. Cardoso and S. Yoshida, JHEP 0507:009 (2005).

\bibitem{HodHod} S. Hod and O. Hod, Phys. Rev. D {\bf 81}, Rapid communication 061502
(2010) [arXiv:0910.0734].

\bibitem{Beyer2} H. R. Beyer, J. Math. Phys. {\bf 52}, 102502 (2011).

\bibitem{HodN} S. Hod, Phys. Lett. B {\bf 708}, 320 (2012).

\bibitem{Bekch} J. D. Bekenstein, Phys. Rev. D {\bf 7}, 949 (1973).

\bibitem{NoteQ} We shall assume, without loss of generality, that
$Q\geq 0$.

\bibitem{HodPirmass} See S. Hod and T. Piran, Phys. Rev. Lett. {\bf 81}, 1554
(1998) [arXiv:gr-qc/9803004] and references therein.

\bibitem{Monf1} V. Moncrief, Phys. Rev. D {\bf 9}, 2707 (1974).

\bibitem{Monf2} V. Moncrief, Phys. Rev. D {\bf 10}, 1057 (1974).

\bibitem{Hodn} S. Hod, Phys. Lett. B {\bf 713}, 505 (2012).

\bibitem{HodPirpam} S. Hod and T. Piran, Phys. Rev. D {\bf 58},
024017 (1998) [arXiv:gr-qc/9712041]; S. Hod and T. Piran, Phys. Rev.
D {\bf 58}, 024018 (1998) [arXiv:gr-qc/9801001]; S. Hod and T.
Piran, Phys. Rev. D {\bf 58}, 024019 (1998) [arXiv:gr-qc/9801060].

\bibitem{Stro} T. Hartman, W. Song, and A. Strominger, JHEP 1003:118 (2010).

\bibitem{HodCQG2} S. Hod, Class. Quant. Grav. {\bf 23}, L23 (2006) [arXiv:gr-qc/0511047].

\bibitem{Hodpla} S. Hod, Phys. Lett. A {\bf 374}, 2901 (2010)
[arXiv:1006.4439].

\bibitem{Hartle} J. B. Hartle and D. C. Wilkins, Commun. Math. Phys. {\bf 38}, 47 (1974).

\bibitem{Heun} A. Ronveaux, {\it Heun's differential equations}.
(Oxford University Press, Oxford, UK, 1995).

\bibitem{Flam} C. Flammer, {\it Spheroidal Wave Functions} (Stanford
University Press, Stanford, 1957).

\bibitem{Abram} M. Abramowitz and I. A. Stegun, {\it Handbook of
Mathematical Functions} (Dover Publications, New York, 1970).

\bibitem{Will} L. E. Simone and C. M. Will, Class. Quantum Grav. {\bf 9}, 963 (1992); R. A. Konoplya and A. Zhidenko, Phys. Rev. D {\bf 73}, 124040
(2006); Rev. Mod. Phys. {\bf 83}, 793 (2011); S. Hod, Phys. Rev. D
{\bf 84}, 044046 (2011) [arXiv:1109.4080]; Y. D\'ecanini, A.
Folacci, and B. Raffaelli, Phys. Rev. D {\bf 84}, 084035 (2011).




\end{thebibliography}
\end{document}